\documentclass[conference]{IEEEtran}

\input epsf
\usepackage{graphicx}
\usepackage[caption=false]{subfig}
\usepackage{algpseudocode,algorithm,algorithmicx}
\usepackage{multirow}
\usepackage{hhline}
\usepackage{diagbox}

\begin{document}

\title{Trua: Efficient Task Replication for Flexible User-defined Availability in Scientific Grids}

\author{\authorblockN{Zhe Zhang\IEEEauthorrefmark{1}, Brian Bockelman\IEEEauthorrefmark{2}, Derek Weitzel\IEEEauthorrefmark{1}, Xinkai Zhang\IEEEauthorrefmark{3}, Hamid Vakilzadian\IEEEauthorrefmark{3}, David Swanson\IEEEauthorrefmark{1}\\ zhan0915@huskers.unl.edu, bbockelman@morgridge.org, dweitzel@cse.unl.edu\\ xinkai.zhang@huskers.unl.edu, hvakilzadian@unl.edu, dswanson@cse.unl.edu}
\authorblockA{\IEEEauthorrefmark{1}Holland Computing Center, University of Nebraska $-$ Lincoln, Lincoln, NE 68588, USA}
\authorblockA{\IEEEauthorrefmark{2}Morgridge Institute for Research, Madison, WI 53715, USA}
\authorblockA{\IEEEauthorrefmark{3}Department of Electrical Engineering, University of Nebraska $-$ Lincoln, Lincoln, NE 68588, USA}
}

\maketitle

\begin{abstract}

Failure is inevitable in scientific computing. As scientific applications and facilities increase their scales over the last decades, finding the root cause of a failure can be very complex or at times nearly impossible. Different scientific computing customers have varying availability demands as well as a diverse willingness to pay for availability. In contrast to existing solutions that try to provide higher and higher availability in scientific grids, we propose a model called Task Replication for User-defined Availability (Trua). Trua provides flexible, user-defined, availability in scientific grids, allowing customers to express their desire for availability to computational providers. Trua differs from existing task replication approaches in two folds. First, it relies on the historic failure information collected from the virtual layer of the scientific grids. The reliability model for the failures can be represented with a bimodal Johnson distribution which is different from any existing distributions. Second, it adopts an anomaly detector to filter out anomalous failures; it additionally adopts novel selection algorithms to mitigate the effects of temporary and spatial correlations of the failures without knowing the root cause of the failures. We apply the Trua on real-world traces collected from the Open Science Grid (OSG). Our results show that the Trua can successfully meet user-defined availability demands.

\end{abstract}
\IEEEoverridecommandlockouts

\IEEEpeerreviewmaketitle

\section{Introduction}
\label{sec:intro}

Understanding and overcoming failures in computing systems have always been an arduous challenge. As geo-distributed cyber-infrastructure emerges for the last decades, these failures face tremendous growth in both scale and complexity. As a result, failure analysis in such systems has become extremely complicated due to the deployment of hundreds of thousands of machines with complex interactions.

Many researchers have studied failure along with its sources and consequences in large-scale computing infrastructures \cite{failure1}\cite{failure2}. Failure in such systems can have major negative consequences such as propagated service disruptions \cite{failure4}, considerable energy waste \cite{failure5}, and significant delay of job completion \cite{failure6}. Although scientific computing providers have used many effective techniques to increase the reliability of their infrastructures, to date, failures are still frequent. In addition, failures in scientific computing systems are not isolated events. Many studies have discovered temporal and spatial correlations among system failures \cite{temporalspatial1}\cite{temporalspatial2}. 

Large scientific computing providers offer multiple levels of Quality-of-Service (QoS) to deal with workload heterogeneity \cite{qos}. Task availability is one of the most important QoS in scientific computing. A priority-based scheduling policy is commonly used to guarantee that service requests submitted to the different QoS levels achieve the desired availability. If needed, resources servicing lower priority requests can be preempted to allow the service of higher priority ones. In this context, tasks running during resource contention periods may be preempted by higher priority tasks, which results in task failure. In preemption-allowed scientific grid systems, preemptions can cause more than 40\% of tasks failed \cite{Zhang:2018:DJP:3219104.3229282}, which makes the task failures as the norm instead of exceptions.

In order to meet user-defined availability, task replication is a commonly used mechanism in scientific grids. Replication is based on an assumption that the probability of a single resource failure is much higher than that of simultaneous failures of multiple resources. It avoids task re-computation by starting several copies of the same task on different resources. With redundant copies of a job, a grid system can continue to provide service in spite of the failure of some grid resources.

In \cite{replication1}, the authors presented a distributed fault-tolerant scheduling algorithm that couples task scheduling with task replication. Their algorithm is static because it depends on using a fixed number of replicas for each task. One of the main disadvantages of static replication is excessively utilizing resources in grids. To overcome this, adaptive replication was proposed \cite{repspatial1}\cite{repspatial2}\cite{repspatial3}\cite{Litke:2007:ETR:1232286.1232287}. In general, adaptive replication proactively creates a fault-tolerant scheduling system by estimating the fault rate on different sites. The number of replicas is dynamic and is determined according to the fault rate of resources scheduled for tasks. Compared with static replication, adaptive replication can save up to 60\% of resource utilization when a grid system runs heavy workload \cite{repspatial2}.

One important domain that the existing research is still missing is the virtualization technique that has been broadly adopted into modern scientific grids. A common reliability model used in adaptive replication is a Weibull distribution \cite{weibullmodel}, which is validated in hardware-based grid systems \cite{weibull1}\cite{weibull2}\cite{weibull3}. However, in virtualized scientific grids \cite{Turilli:2018:CPP:3186333.3177851}\cite{glideinWMS}, reliability model might not be able to be estimated by a Weibull distribution \cite{vm1}. For example, the failure rate generated from a Weibull distribution is monotonically decreasing as the machine prolongs its lifetime \cite{Litke:2007:ETR:1232286.1232287}. However, in modern scientific grids such as the OSG, a pilot (similar to a virtual machine in cloud computing) is originated with a deadline which complies with the estimated runtime of the task which is scheduled to run on the pilot. The task will be immediately killed after the pilot passes its deadline, which means the pilot is saturated at the maximum failure rate as it passes the deadline. In other words, the failure rate is impossible to infinitely decrease. In addition to the reliability model, the spatial correlation of failures in the system can also be changed. A virtual layer introduced in the scientific grid underpins the heterogeneous hardware with homogeneous pilots and expose a universal interface to the users. Physically localized failures are not necessary to generate task failures in the virtual layer.

In this paper, we investigate the task failures in the OSG which is one of the state-of-the-art scientific grids. This paper has two contributions to the community of scientific computing:

\begin{itemize}
\item
We found the reliability model of a pilot lifetime in the OSG can be represented by a bimodal Johnson distribution, which is different from any existing reliability models in the field.
\item
Based on the reliability model, we proposed the Trua. Trua uses an anomaly detector to filter out anomalous failures (temporal and spatial correlated failures) in the system. Trua also uses novel selection algorithms to deploy the replicas to the adequate pilots in which the user-defined task availability can be met.
\end{itemize}

In this paper, we use the term - \textit{pilot} - to represent a virtual instance in scientific grids. A task submitted to the grids will be eventually executed on a pilot. This concept is similar to running a task on a virtual machine in the cloud environments. Another important term used in this paper is \textit{failure rate}. In reliability engineering, failure rate corresponds to the inverse of mean time to failure. We comply with this concept for measuring the lifetime statistics of pilots in Section \ref{sec:bathtubcurve}. However, in the rest of the paper, we use it to describe task failures. It corresponds to a complement of task availability.
\section{Related Work}
\label{sec:related}

Traditional adaptive replication consists of two models - a failure model and a replication model. Since Litke \cite{Litke:2007:ETR:1232286.1232287} provides a general representation of the adaptive replication, in this paper, we call them together as Litke's model. Let us review the formal definition of these two models in this section.

\subsection{Failure Model}
\label{sec:failuremodel}

The first model focuses on running a single copy of a task on a pilot. It aims to quantify a task's failure probability based on a pilot lifetime PDF. Without losing any generality, we assume that a task is going to start its execution at time $t_0$. This also implies a pilot which carries the task is alive at time $t_0$. The failure rate of the task, as expressed in Equation \ref{eq:section4:eq5}, is defined as the probability of failure during the next $\Delta t$ time units. 

\begin{equation}
\label{eq:section4:eq5}
P(t_0<s<t_0+\Delta t|s>t_0) = P(t_0<s<t_0+\Delta t)/P(s>t_0)
\end{equation}

where $s$ represents the time that the task fails to continue to run.

We use $p(t)$ to represent the runtime PDF of the pilot job and $f_i$ to represent the failure rate of the task $T_i$, then the above equation can be further represented as follows:

\begin{equation}
f_i = \int_{t_0}^{t_0+\Delta t}p(t) / \int_{t_0}^{\infty}p(t)
\end{equation}

\subsection{Replication Model}
\label{sec:replication}

Litke's model then decides to use replication to protect a task from failures. The term replica is used to denote an identical copy of a task. By producing task replicas, the failure probability for a task can be significantly lowered. We assume that $m_i$ replicas - denoted by $T_{ik}$, $k = 1,...,m_i$ - of a task $T_i$ are produced and placed in the grid. The new failure rate for task $T_i$ becomes:

\begin{equation}
f_i' = \prod_{k=1}^{m_i} f_{ik}
\end{equation}

The above equation corresponds to the probability of the event ``all replicas fail (the original task is also considered as a replica).'' The number of replicas created depends on the failure probabilities of replicas and the desired availability level from the application. We denote $\lambda$ as the desired availability. Thus, for each task, a sufficient number of replicas should satisfy the condition:

\begin{equation}
\prod_{k=1}^{m_i} f_{ik} \le 1-\lambda
\end{equation}

where $\lambda \in (0,1)$.

A loss of a single replica should not be considered as a failure because other replicas might survive to the end. We consider a task successfully finishes when any of the replicas finish in the end.

From the OSG perspective, for any user-submitted task, there are two parameters attached to it: target availability $\lambda$ and lease period $\Delta t$. A lease period is the estimated runtime for a task to run on a pilot. If the required lease period is larger than the assigned pilot's lifetime, the task cannot finish before the pilot terminates and therefore the task will fail. The OSG needs to look at the pilot pool and select a set of pilots so that their cumulative availability is greater than the target availability. More importantly, if we sort all pilots by their failure rates and select pilots from the most reliable one, Litke's model can guarantee that we use the minimum redundancy to replicate a task for any given availability \cite{Litke:2007:ETR:1232286.1232287}. In this paper, we evaluate both the random selection and the sorted selection and use Random and Sorted to represent two selections respectively.
\section{Exploration}
\label{sec:exploration}

This section is to build a case study for Litke's model. We monitored the OSG's pilots for 102 days and collected their running information such as start time, end time and so on. We additionally collected 34 days of pilot data for the test purpose. We first characterize the pilot lifetime PDF. Based on that, we test the model. The results show that the model barely achieves the target availability levels due to the large variance in the pilot lifetime PDF and the deficiency of the model itself.

\subsection{Lifetime PDF}
\label{sec:pdf}

\begin{figure}[!ht]
\begin{center}
\includegraphics[height=2.5in, width=3.4in]{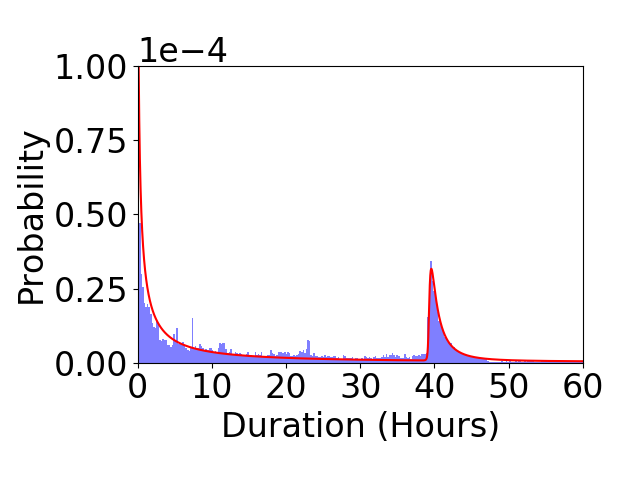}
\end{center}
\vspace*{-5mm}
\caption{Pilot lifetime PDF}
\label{fig:estimationpdfsuogce1}
\end{figure}

Figure \ref{fig:estimationpdfsuogce1} shows the lifetime PDF on 102-day dataset. Unlike Litke's assumption that the lifetime PDF follows a Weibull distribution, the actual PDF follows a bimodal Johnson distribution. In our previous work \cite{Zhang:2018:DJP:3219104.3229282}, we divided pilot termination into two categories: expected termination and unexpected termination and use two Johnson distributions to fit two PDFs. In order to understand the difference between expected and unexpected terminations, we need to introduce some background knowledge. In the OSG, there are two important parameters pertaining to pilot lifetime - retire time and kill time. Retire time is considered as a ``soft deadline" for a pilot and kill time is considered as a ``hard deadline". After passing the retire time, a pilot is still able to run its carried task until the task is finished. If a carried task cannot be finished before its carrier pilot reaches the kill time, the pilot will be intermediately killed by the system resulting in task failure. Retire time and kill time are configurable in the OSG, they can be adjusted based on different workloads. During the time period we collected the data, all pilots' retire time and kill time had been configured around 38 hours and 40 hours. In Figure \ref{fig:estimationpdfsuogce1}, all pilots whose lifetime were less than the retire time were terminated due to unexpected failures such as preemptions and network disconnections; the rest of pilots whose lifetime greater than the retire time were terminated due to normal pilot life cycle.

As described in our previous work \cite{Zhang:2018:DJP:3219104.3229282}, preemptions are apt to happen in an early stage. In Figure \ref{fig:estimationpdfsuogce1}, most of the pilots whose lifetimes are less than 10 hours are terminated due to preemptions. Preemptions are dominant in unexpected terminations and leave other unexpected failure types (network disconnections, idle shutdown and so on) negligible. We also call this region as the preempting stage. After the preempting stage, a pilot enters a stable stage where pilots have less chance to be terminated. The stable stage lasts until the retire time is reached and from the moment a pilot enters the retiring stage.

\subsection{Bathtub Curve}
\label{sec:bathtubcurve}

\begin{figure}[!ht]
\begin{center}
\includegraphics[height=2in, width=3in]{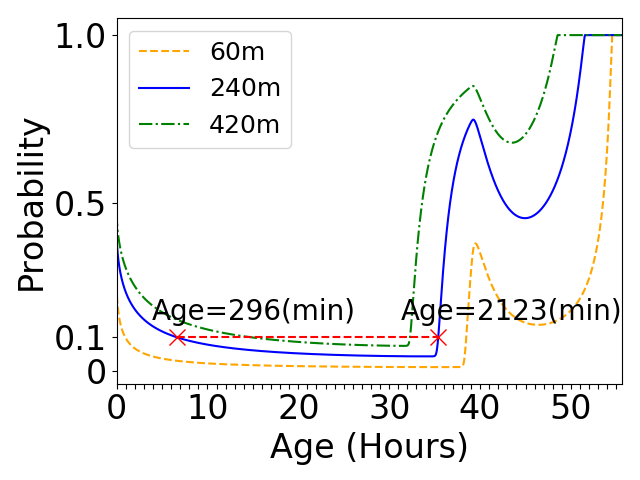}
\end{center}
\vspace*{-5mm}
\caption{This figure shows failure rate curves based on three lease periods: 60, 120, 240 minutes; this figure also indicates a failure rate can be ambiguously interpreted into two pilot running ages}
\label{fig:ambiguity}
\end{figure}

We applied Litke's failure model to the lifetime PDF and tested on three lease periods - 60, 240, 420 minutes. Figure \ref{fig:ambiguity} shows three failure rate curves generated from Figure \ref{fig:estimationpdfsuogce1}. The failure rate curves demonstrate three stages in a pilot life cycle: in the preempting stage, a pilot faces high failure rate; as a pilot survives preemptions, it goes into the stable stage and its failure rate keeps decreasing as the age increases; however, when a pilot approaches the retire time, the failure rate starts sharply increasing.

\begin{figure}[!ht]
\begin{center}
\includegraphics[height=1.2in, width=2.5in]{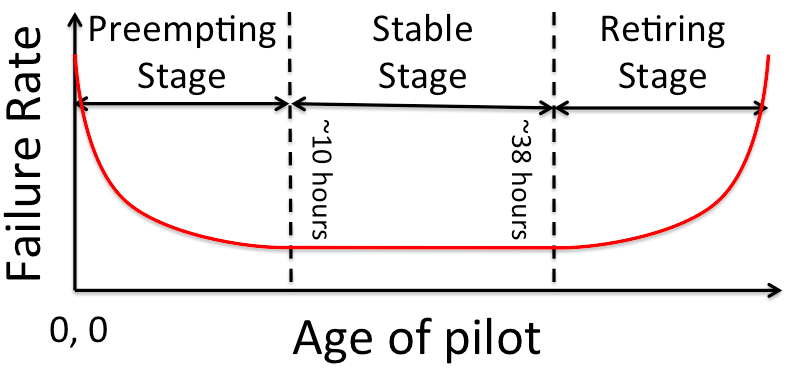}
\end{center}
\vspace*{-5mm}
\caption{A bathtub curve}
\label{fig:bathtubcurve}
\end{figure}

According to Litke's replication model, a task can produce a large number of replicas if the selected pilots have high failure rates, which results in wasting resources. On a failure rate curve, left region before retire time is more attractive to our design. We call such a truncated region as a bathtub curve and Figure \ref{fig:bathtubcurve} conceptually summarizes a bathtub curve pertaining to the OSG pilots.

\subsection{Pitfalls of the Model}
\label{sec:pitfalls}

We tested Litke's model on our 102-day pilot dataset. Since scanning the whole dataset takes too long, we sampled our dataset every 100 minutes. At each sample, we listed all available pilots and calculated their ages. With Litke's failure model, we can estimate each pilot's failure rate. After getting all failure rates, we run Litke's replication model. We conducted two algorithms on selecting pilots: random selection and sorted selection. Both selections accumulate multiple pilots until their overall failure rate reaches above a target availability. The difference is that random selection randomly picks pilots and sorted selection always starts selecting from the most reliable one.

\begin{figure}[!ht]
\begin{center}
\includegraphics[height=2.5in, width=3.2in]{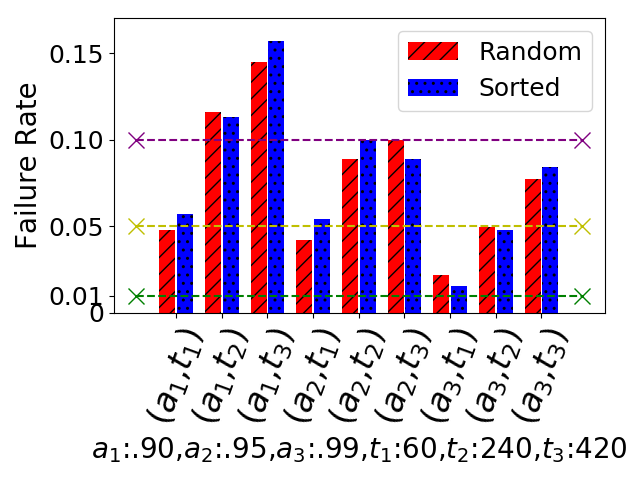}
\end{center}
\vspace*{-5mm}
\caption{The first three pairs of histograms which set target availability as 0.90 are expected to have failure rate below 0.10; the middle three pairs of histograms are expected to be below 0.05; the last three pairs are expected to be below 0.01.}
\label{fig:explorationfr}
\end{figure}

Figure \ref{fig:explorationfr} shows the overall failure rates on different combinations of target availability levels and lease periods. Each combination is represented by a pair ($a(n)$, $t(n)$) where $a(n)$ represents the availability levels (0.90, 0.95, 0.99) and $t(n)$ represents the lease periods (60, 240, 420 minutes). As shown in the figures, only the test of (0.90, 60) which is represented by ($a_1$, $t_1$) meets the target availability. For the test of (0.95, 60) which is represented by ($a_2$, $t_1$), random selection meets the target availability; however, sorted selection fails to keep the failure rate under 0.05. In general, the model gets closer to the target availability at a shorter lease period (60 minutes). As the lease period becomes larger, the model becomes more inaccurate to select reliable pilots.

\begin{figure}[!ht]
\begin{center}
\includegraphics[height=2.5in, width=3.2in]{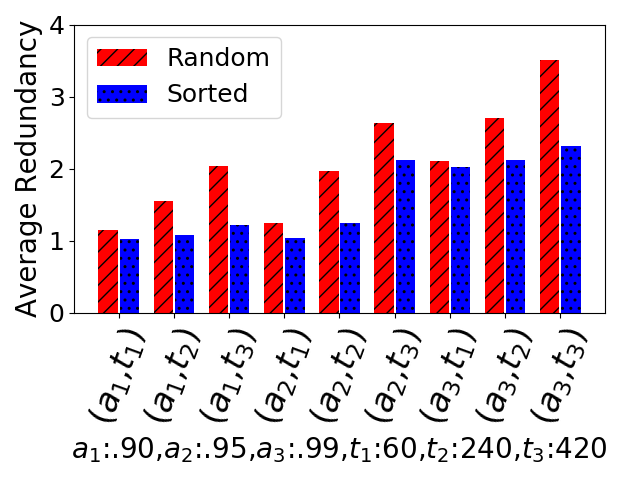}
\end{center}
\vspace*{-5mm}
\caption{Redundancy costs of Litke's model}
\label{fig:explorationredundancy}
\end{figure}

Figure \ref{fig:explorationredundancy} shows the average number of replicas that are selected for a task. As the sorted algorithm always selects the most reliable pilots, it uses the minimum redundancy. Interestingly, if we look at both Figure \ref{fig:explorationredundancy} and Figure \ref{fig:explorationfr}, the tests of (0.90, 240), (0.95, 420), (0.99, 60) and (0.99, 240) show that sorted selection uses less redundancy to achieve higher availability compared with random selection. This implies sorting failure rate indeed helps to improve task availability. Unfortunately, neither of the two algorithms can meet the target availability levels.

\section{Analysis}
\label{sec:analysis}

We summarize the reasons that make Litke's model fail to select adequate pilots as follows.

\subsection{Ambiguity in a bathtub curve}
\label{sec:ambiguity}

The Weibull distribution in Litke's assumption generates a monotonically increasing failure rate curve. However, a bimodal Johnson distribution generates a non-monotonical curve.  As two red x-points show in Figure \ref{fig:ambiguity}, two pilots who run into different ages can have an equal failure rate. In such a case, one pilot is still in its early stage and the other is in the retiring stage. A user might prefer to select the first pilot because as it continues to run, its failure rate will decrease. On the other hand, the second pilot's failure rate will sharply increase. For this reason, the model might not be able to select a set of the most reliable pilots.

\subsection{Anomoly in pilot lifetime PDF}
\label{sec:anomalyissue}

\begin{figure}[ht!]
     \centering
     \subfloat[An anomaly in pilot lifetime PDF]{\includegraphics[height=1.4in, width=1.7in]{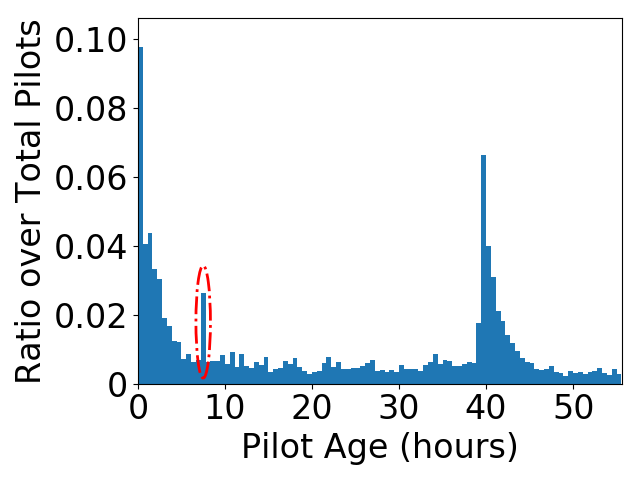}\label{fig:anomaly1}}
     \subfloat[Network failure distribution]{\includegraphics[height=1.4in, width=1.7in]{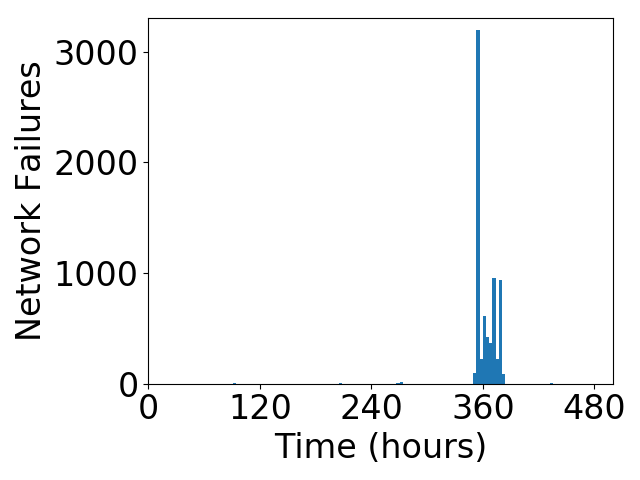}\label{fig:anomalynetworkissue}}\\
     \subfloat[Pilot start time distribution of the pilots that have lifetime between 7 and 8 hours]{\includegraphics[height=1.4in, width=1.7in]{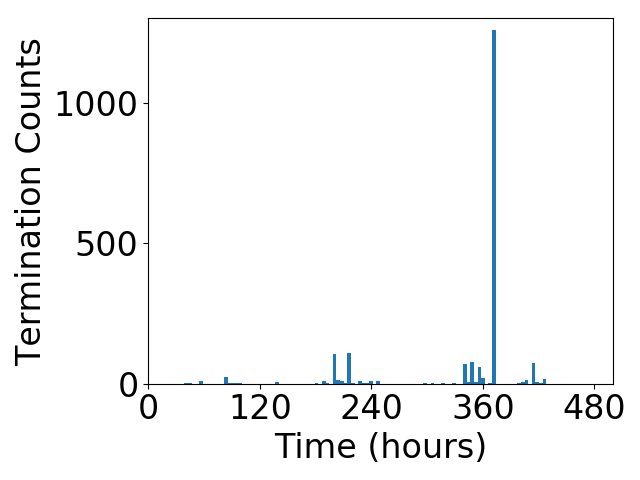}\label{fig:anomalystarttime}}
     \subfloat[Zoom-in of start time]{\includegraphics[height=1.4in, width=1.7in]{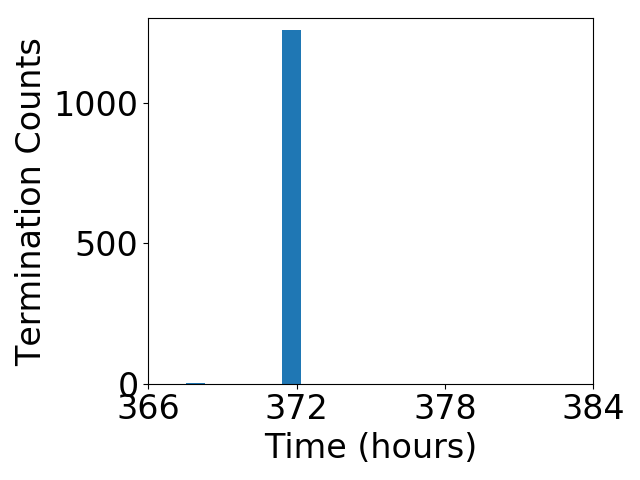}\label{fig:anomalystarttimezoomin}}\\
     \subfloat[Pilot end time distribution of the pilots that have lifetime between 7 and 8 hours]{\includegraphics[height=1.4in, width=1.7in]{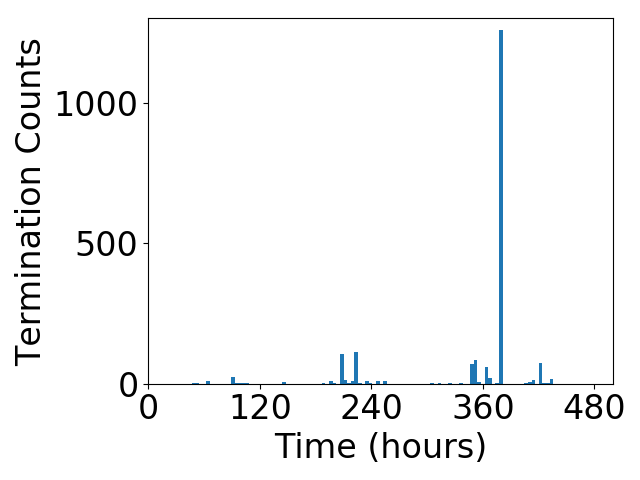}\label{fig:anomalyendtime}}
     \subfloat[Zoom-in of end time]{\includegraphics[height=1.4in, width=1.7in]{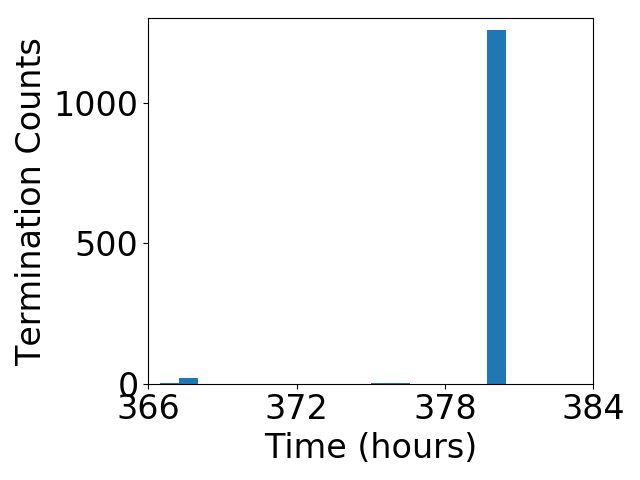}\label{fig:anomalyendtimezoomin}}
     \caption{Anomaly in the pilot lifetime distribution}
     \label{fig:anomalytime}
\end{figure}

Incidents such as power loss or network disconnection can cause a large number of failures resulting in a sudden rise in pilot terminations. We call such unexpected pilot terminations as anomalies. To demonstrate anomalies in the OSG, we select the first 18 days in the 102-day pilot dataset where a large number of pilot failures occurred due to network disconnection. Figure \ref{fig:anomaly1} shows the lifetime PDF of the 18-day data. A red circle indicates a large spike in the lifetime PDF that is about 7 hours to 8 hours. We extract all pilots with a lifetime between 7 and 8 hours and examine their pilot start time and end time. Figure \ref{fig:anomalystarttime} and \ref{fig:anomalystarttimezoomin} show start time distribution of those pilots. Figure \ref{fig:anomalyendtime} and \ref{fig:anomalyendtimezoomin} show end-time distribution of those pilots. These figures show a large number of pilots start and end around 370 and 380 hours. After adopting the pilot classifier described in our previous work \cite{Zhang:2018:DJP:3219104.3229282}. It turns out a large number of network disconnections happened in the OSG. Figure \ref{fig:anomalynetworkissue} shows that we detected a sudden rise of network failures in our pilot dataset. 

\subsection{Variance in pilot lifetime PDF}
\label{sec:randomness}

Another factor that makes the model difficult to select the most reliable pilots is the variance in the lifetime PDF caused by non-anomalous preemptions in the OSG. In our previous work, we found that over 40\% of pilots encountered preemptions. Most of those preemptions happened on a regular basis and bring the randomness in pilot lifetime PDF overall time ranges. In addition, the OSG usually disassembles a large job into multiple tasks in order to execute them in parallel. As a result, the OSG has to create a bunch of pilots around the same time to carry those tasks if there are not enough pilots available in the pilot pool. These newly created pilots have the same start time and therefore will have the same end time if preemptions happened to these pilots. We call this effect as \textit{lifetime locality}. In Litke's model with failure rate being sorted, these pilots are highly likely to be selected together because they have similar age and therefore similar failure rate. This pilot behavior has been discussed in more detail in our previous work \cite{Zhang:2018:DJP:3219104.3229282}, in this paper we will propose new pilot selection algorithms in Section \ref{sec:methodology} to specifically tackle this effect.
\section{Architecture of the Trua}
\label{sec:architecture}

\begin{figure}[!ht]
\begin{center}
\includegraphics[height=2.2in, width=3.4in]{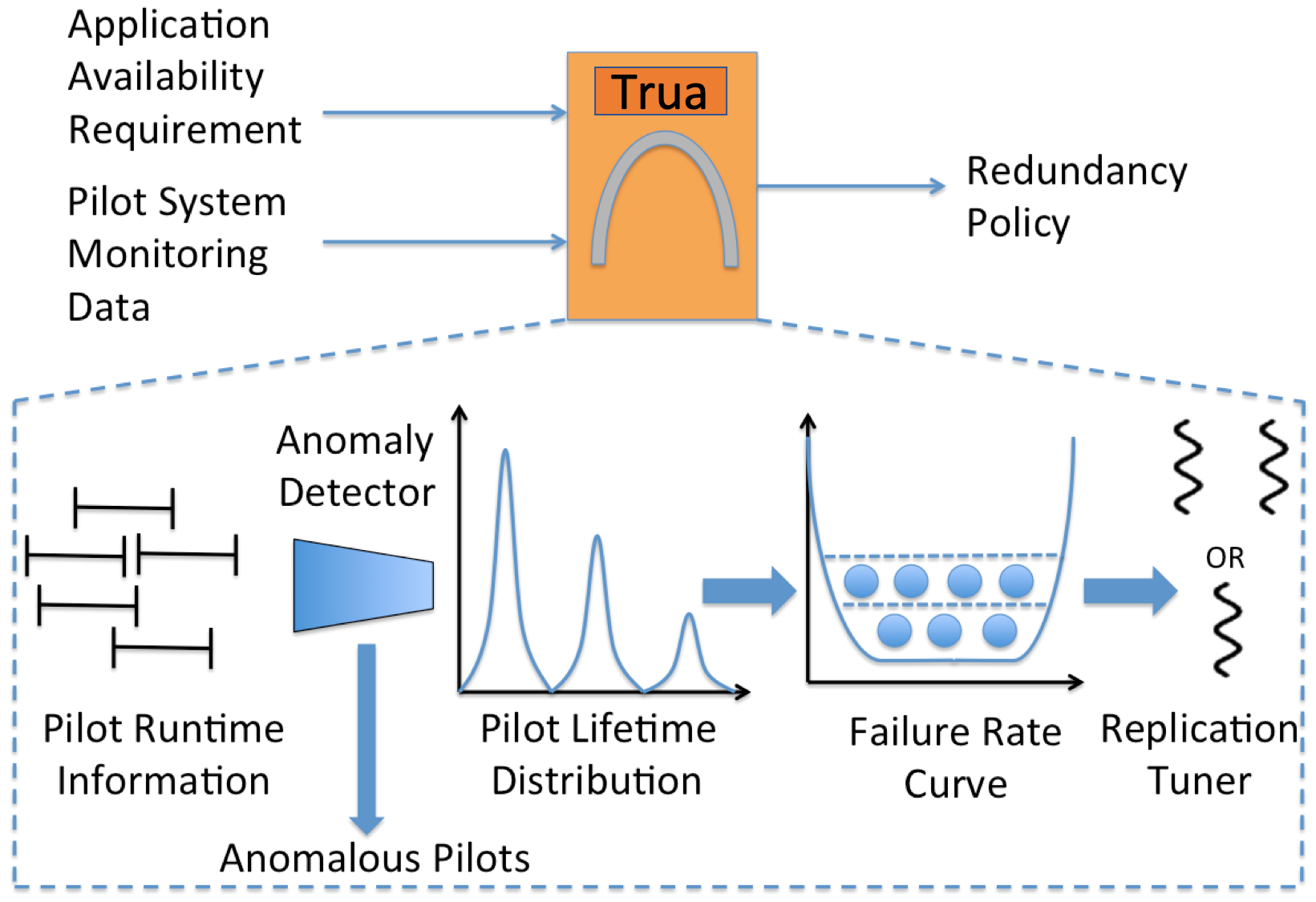}
\end{center}
\vspace*{-5mm}
\caption{System architecture of Trua}
\label{fig:architecture}
\end{figure}

Figure \ref{fig:architecture} shows the workflow of the Trua. It adopts an anomaly detector to monitor pilot runtime information. If any suspicious pilot failures arise, the Trua can filter out the infected pilots. Following an anomaly check, a pilot lifetime distribution can be generated from the collected information and consequently multiple bathtub curves, each of which pertains to a lease period, can be derived from the lifetime distribution. Finally, depending on target availability, the Trua tunes redundancy assigned for a task and make sure selected pilot replicas can meet target availability.

\subsection{Anomaly Detection}
\label{sec:anomalydetection}

\begin{figure}[ht!]
     \centering
     \subfloat[Before Removing Anomalies]{\includegraphics[height=1.4in, width=1.85in]{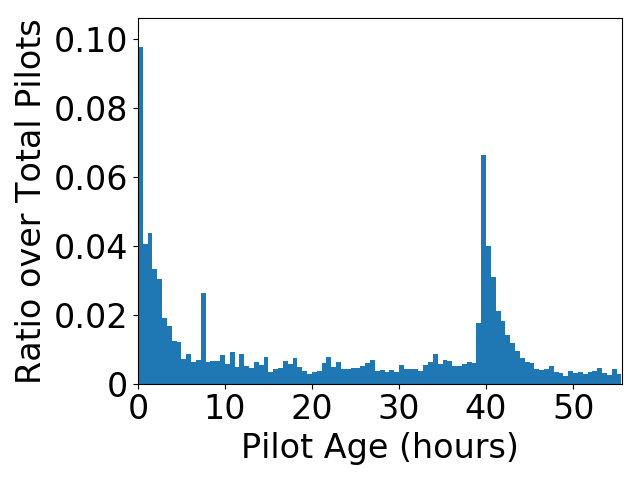}\label{fig:anomalybefore}}
     \subfloat[After Removing Anomalies]{\includegraphics[height=1.4in, width=1.55in]{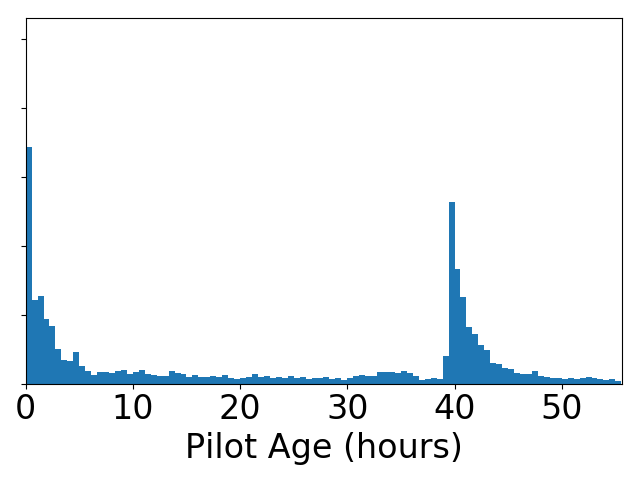}\label{fig:anomalyremoval}}
     \caption{Effect of anomaly detector}
     \label{fig:anomalydetector}
\end{figure}

In order to detect failure anomalies, we adopt the RRCF algorithm \cite{Guha:2016:RRC:3045390.3045676}. We put the pilots that are characterized as network failures in a time stream. The anomaly detector acts on the time stream. The output from the anomaly detector is also a data stream containing anomaly scores produced by RRCF algorithm. Potential anomalies identified by RRCF have a higher anomaly score than data that the algorithm considers non-anomalous. RRCF generates the anomaly score based on how different the new data is compared to the recent past. If the anomaly score is above a certain threshold, the detector considers the recently terminated pilots are anomalous. Figure \ref{fig:anomalyremoval} shows the lifetime PDF after removing anomalies. 

The anomalous failures can easily exceed the limits of any reasonable replication scheme, so we should not replicate tasks when an anomaly appears. Instead, the OSG should halt scheduling tasks to any infected cluster sites until the incidents get resolved. Therefore, we select another input parameter in our anomaly detector which defines a halting period after detecting an anomaly. Figure \ref{fig:anomalyremoval} shows the new pilot lifetime PDF after we applied RRCF with a threshold score set to 250 and a halting period set to 15 minutes.

\subsection{Valleys in a Bathtub Curve}
\label{sec:valleyidea}

\begin{figure}[!ht]
\begin{center}
\includegraphics[height=1.2in, width=2.5in]{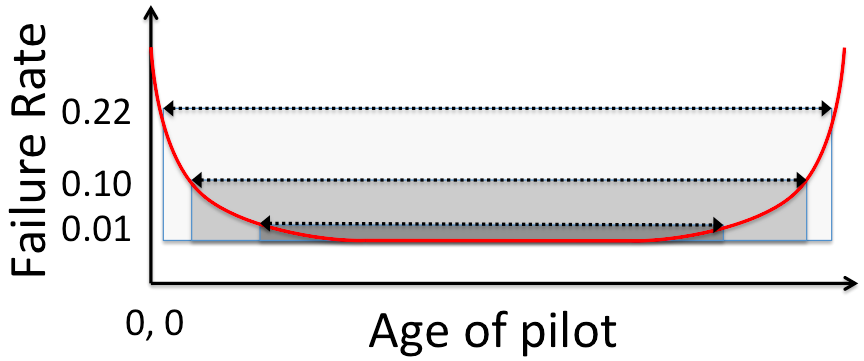}
\end{center}
\vspace*{-5mm}
\caption{Valleys in a bathtub curve}
\label{fig:bathtubvalleys}
\end{figure}

Although the anomaly detector can filter out anomalous preemptions, most of the preemptions in the OSG are non-anomalous. In order to overcome these failures, we introduce a concept - \textit{valley}. Figure \ref{fig:bathtubvalleys} shows three valleys in a bathtub curve. A valley is defined by a target availability (failure rate is (1 - availability)). A smaller valley covers a shorter lifetime range. Therefore, the number of pilots that exist in a smaller valley is less than the number of pilots in a larger valley. More importantly, all those pilots that exist in a smaller valley are always a subset of the pilots that exist in a larger valley. With the definition of the valley, let us look at why we want to determine valleys in a bathtub curve.

\subsubsection{Overcome variance in the PDF}

Valleys can overcome the variance in a pilot lifetime PDF. If we can determine a stable valley that the pilots within the valley have high reliability, then the pilots can be randomly selected from the valley. The selected pilots are expected to meet target availability. Due to lifetime locality, it is commonly seen in the OSG that a group of pilots start and end together. If any pilot in a group gets selected, sorted selection in Litke's model highly likely selects other pilots in the group as well. If those pilots face preemptions, the subsequent replicas cannot do any favor to improve the availability. Random selection in a valley can definitely overcome this issue. In addition to lowering failure rate, random selection can also improve system utilization. Intuitively, the model in Section \ref{sec:related} can result in load unbalance in the system by assigning tasks to the most reliable pilots which might only consist of a small portion of available pilots.

\subsubsection{Keep task redundancy to the minimum}

As shown in the sorted selection of Section \ref{sec:replication}, the algorithm immediately stops as the cumulative failure rate of the most reliable pilots exceeds the availability requirement. The algorithm can assure that the system creates the minimum number of task replicas to meet the target availability.

Valleys can also keep the redundancy to the minimum by carefully defining valleys by certain availability levels. For example, if the target availability is 0.99, we can choose any pilot within the valley region under failure rate 0.01 to achieve it. However, high-availability valleys contain fewer pilots than low-availability valleys. As a result, a smaller valley might not have enough pilots to meet target availability. To address this problem, we should consider multiple valleys for target availability. As shown in Figure \ref{fig:bathtubvalleys}, availability of 0.99 can be met with one pilot within the valley 0.01, or two pilots within the valley 0.1 ($\sqrt[2]{0.01} = 0.1$), or three pilots within the valley 0.22 ($\sqrt[3]{0.01} \approx 0.22$), and even more pilots on larger valleys (not shown in the figure). Compared with the sorted selection, randomly selecting pilots from valleys should not increase the task redundancy in the system.
\section{Methodology}
\label{sec:methodology}

\subsection{Challenges}
\label{sec:challenges}

There are several challenges in practically adopting the idea of using valleys to our system.

\subsubsection{Does pilot lifetime distribution change quickly over time?}

The Trua relies on the historical pilot runtime information to generate the lifetime PDF. If the pilot lifetime distribution varies frequently over time, the estimated lifetime PDF and valleys might not be adequate to represent real-time pilot characteristics.

\subsubsection{How valleys can be accurately determined?}

As shown in Section \ref{sec:pitfalls}, due to the variance in a pilot lifetime, the failure rate curve cannot 100\% accurately estimate the actual reliability. We need to make sure the estimation error is in a reasonable range.

\subsection{Generating Lifetime PDF}
\label{sec:framepdf}

We use our data collected from the OSG over 102 days. We split them into three time frames, each of which contains ~ 80,000 pilots. We apply the anomaly detector with the parameters - score: 250, delay: 60 minutes - to the three time frames.

\begin{figure}[!ht]
\begin{center}
\includegraphics[height=2in, width=3in]{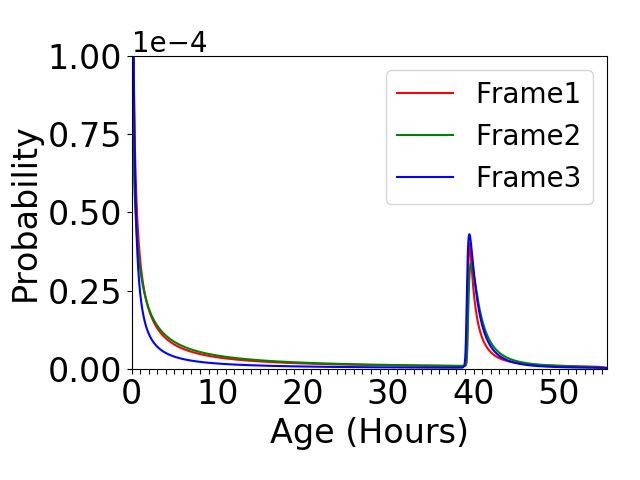}
\end{center}
\vspace*{-5mm}
\caption{Pilot lifetime PDFs of three time frames}
\label{fig:comparepdf}
\end{figure}

Figure \ref{fig:comparepdf} shows the PDF fitting curves on the three time frames. As seen in the figure, the three time frames follow the same shape. Although anomalies unpredictably happen over time, once we group enough number of pilots, the pilot lifetime PDF is relatively stationary to some degree. This observation confirms the feasibility of using historical pilot information to estimate the ongoing pilot failure rate. We do not run a sensitivity analysis on the size of historical pilots that are eligible to estimate the pilot lifetime PDF. More investigations might be necessary to identify the minimum time window or the size of pilots. 

\subsection{Determining Valleys on Failure Rate Curves}
\label{sec:curve}

We use a statistical approach to generate failure curves given a set of pilots. The procedure works as follows.

\subsubsection{Select the granular interval to characterize failure rate}

We select a small size of the time interval. We divide the pilot lifetime into multiple intervals. For example, the pilot lifetime is between 0 and 200,000 seconds. If we select the time interval as 200 minutes (12,000 seconds), there will be 17 intervals in lifetime distribution.

\subsubsection{Calculate failure rate on individual intervals}

Given a lease period, we test each interval overall dataset (102 days) and calculate the failure rate according to each interval. For example, given a lease period of 240 minutes, we first select a pilot only within the age range between 0 and 200 minutes. We test the range (0, 200] over 102 days and calculate the average failure rate for this interval. Then we move the interval to (200, 400] and test the new interval over 102 days and so on. When we finish estimating the failure rate of all intervals, we plot the failure curve. Note we only select one replica in each interval in the aforementioned test. We need to repeat the same test but select a different number of pilots within each interval in order to test failure rate curves for different redundancy levels. We do not need to run the test with a large number of replicas. We should stop moving to a higher redundancy level once the valleys determined by existing redundancy levels can cover the range between 0 and the retire time.

\subsubsection{Determine valleys on failure rate curves}

\begin{figure}[!ht]
\begin{center}
\includegraphics[height=2.5in, width=3.4in]{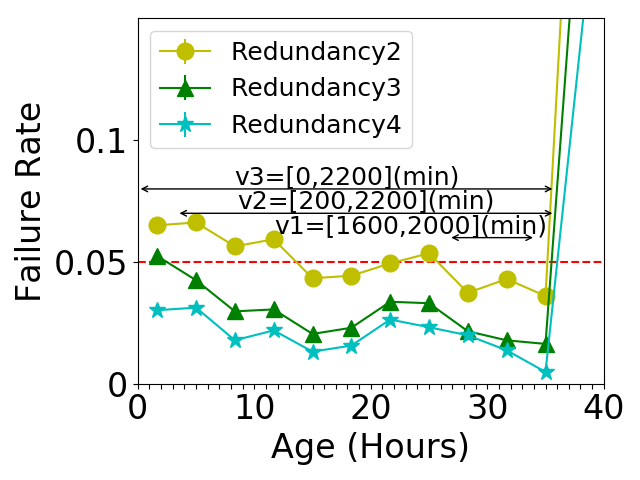}
\end{center}
\vspace*{-5mm}
\caption{Valleys on Failure Curves: Valley1(v1) is [1600,2000] minutes and 2 pilots need to be selected to meet the availability which is 0.95; Valley2(v2) is [200,2200] and 3 pilots need to be selected; Valley3(v3) is [0,2200] and 4 pilots need to be selected.}
\label{fig:valley}
\end{figure}

After generating multiple failure rate curves on redundancy levels, we draw horizontal lines at the target availability levels to determine the valleys. For example, Figure \ref{fig:valley} shows the valleys that are determined by the target availability of 0.95 and the lease period of 240 minutes. There is no failure rate curve for Redundancy1 because a single replica cannot meet the target availability at any intervals. 

As shown in Figure \ref{fig:valley}, when the system tries to replicate a task to multiple pilots. It should first try to select 2 pilots in the most reliable valley - v1. If there are less than 2 pilots in this valley or pilots are not applicable, the system should relax the valley to v2 and randomly select 3 pilots in the valley. If v2 is not applicable again, the system moves the valley to v3 and tries to select 4 pilots.

\subsection{Algorithms}
\label{sec:algorithms}

After characterizing the valleys of different availability levels, we propose two algorithms to select pilots from those valleys: valley selection and spread selection. We will also use Valley and Spread in this paper to represent these two algorithms. Algorithm \ref{alg:bathtubreplication} shows the procedures of two algorithms. The difference between these two algorithms is that valley selection uses random\_select\_n\_pilots which randomly select $r$ pilots out of all available pilot within a valley; spread selection uses spread\_select\_n\_pilots shown in Algorithm \ref{alg:spread} which tries to evenly distribute the selected pilots over their minimum and maximum start times. The motivation behind spread selection is to mitigate the effect coming from lifetime locality.

\begin{algorithm}  
\caption{Bathtub Replication Algorithm - $P$ is the set of all available pilots at a given time; $V$ is the valley list (sorted from the most reliable valley to the most error-prone valley) associated to a given pair of (availability, lease)}  

\label{alg:bathtubreplication} 
  \begin{algorithmic}
    \Procedure{Select\_Pilots}{$P$, $V$}
      \For{$v_i$ in $V$}
        \State $r \gets$ get\_valley\_redundancy($v_i$)
        \State $p_i \gets$ get\_pilots\_in\_valley($v_i$, $P$)
        \State $n \gets$ get\_length($p_i$)
        \If{$n \ge r$}
          \State $p^s_i \gets$ random\_select\_n\_pilots($r$, $p_i$)
          \State \textbf{or}
          \State $p^s_i \gets$ spread\_select\_n\_pilots($r$, $p_i$)          
          \State \textbf{return} $p^s_i$
        \EndIf
      \EndFor
      \State \textbf{return} NoSolution
    \EndProcedure
  \end{algorithmic}  
\end{algorithm}

Both algorithms start from the most reliable valley in which only the minimum number of pilots need to be selected. The algorithms gradually move to a larger valley and try to select more pilots if the preceding smaller valley is not able to select enough pilots. The algorithms return a set of pilots when there is a valley that can select enough pilots that match the valley's redundancy level.

\begin{algorithm}  
\caption{spread\_select\_n\_pilots function - $r$ is the number of pilots we expect to select; $p$ is a set of pilots that exist in a specific valley; get\_spread\_n\_interval($r$, $t_{min}$, $t_{max}$) returns a list of intervals that are evenly distributed over the time span ($t_{min}$, $t_{max}$] where $t_{min}$ is the minimum pilot start time and $t_{max}$ is the maximum pilot start time}  
\label{alg:spread} 
  \begin{algorithmic}
    \Procedure{spread\_select\_n\_pilots}{$r$, $p$}
      \State $p^s$ = set()
      \State $t_{min} \gets$ get\_min\_start\_time($p$)
      \State $t_{max} \gets$ get\_max\_start\_time($p$)
      \State $l_{intv} \gets$ get\_spread\_n\_interval($r$, $t_{min}$, $t_{max}$)
      \State $idx \gets$ 0
      \While{$len(p^s) < r$}
        \State $v_i \gets l_{intv}[idx]$
        \State $p_i \gets$ get\_pilots\_in\_valley($v_i$, $p$)
        \If{$len(p_i) > 0$}
          \State $p^1_i$ = random\_select\_n\_pilots($1$, $p_i$)
          \State $p^s.insert(p^1_i)$
          \State $p.remove(p^1_i)$
        \EndIf
        \State $idx = (idx + 1) \% r$
      \EndWhile
      \State \textbf{return} $p^s$
    \EndProcedure
  \end{algorithmic}  
\end{algorithm}
\section{Evaluation}
\label{sec:evaluation}

We apply the valleys of different pairs of availability and lease periods that are generated from the 102-day pilot dataset to the test dataset that contains 34-day pilot information. We measure the performance of our model in three metrics: failure rate, redundancy and resource utilization and compare the performance with Litke's model.

\subsection{Valleys}

Table \ref{tab:valleys} shows the valleys for nine combinations of availability levels (0.90, 0.95, 0.99) and lease periods (60, 240, 420 minutes). All valleys are determined based on 102-day pilot data. \textbf{R} represents how many replicas a task needs to be replicated on a certain valley. \textbf{v(n)} represents a valley range that corresponds to a redundancy level. For example, let us look at the pair of the availability of 0.95 and the lease of 420 minutes. There is no valley region that can achieve the target availability of 0.95 by selecting one pilot. Therefore, v1 is labeled as N/A. If we further look at v2, it indicates a valley range between 1800 and 1950 minutes, which means by selecting any two pilots whose ages are between 1800 and 1950 minutes, the cumulative failure rate of the selected pilots is expected to be below 0.05 (the target availability is 0.95). In Table \ref{tab:valleys}, we omit v11 in which none of the availability-lease-pairs have a valid valley.

\begin{table*}[ht]
\centering
\caption{Valleys for different redundancy levels}
\label{tab:valleys}
\begin{tabular}{|c|c|c|c|c|c|c|c|c|c|c|c|}
\hline
A,L & v1 & v2 & v3 & v4 & v5 & v6 & v7 & v8 & v9 & v10 & v12\\
\hline
.90,60 & 50,2100 & 0,2100 & N/A & N/A & N/A & N/A & N/A & N/A & N/A & N/A & N/A \\
.95,60 & 100,2100 & 0,2100 & N/A & N/A & N/A & N/A & N/A & N/A & N/A & N/A & N/A \\
.99,60 & N/A & 1250,2000 & 100,2100 & 50,2100 & N/A & N/A & 0,2100 & N/A & N/A & N/A & N/A\\
.90,240 & 450,2100 & 50,2100 & 0,2100 & N/A & N/A & N/A & N/A & N/A & N/A & N/A & N/A\\
.95,240 & 2000,2050 & 550,2100 & 50,2100 & 0,2100 & N/A & N/A & N/A & N/A & N/A & N/A & N/A\\
.99,240 & N/A & N/A & 1700,1850 & N/A & 1300,2100 & 250,2100 & 50,2100 & N/A & 0,2100 & N/A & N/A\\
.90,420 & N/A & 150,1950 & 50,2000 & 0,2050 & N/A & N/A & N/A & N/A & N/A & N/A & N/A\\
.95,420 & N/A & 1800,1950 & 550,1950 & 50,2000 & N/A & 0,2050 & N/A & N/A & N/A & N/A & N/A\\
.99,420 & N/A & N/A & N/A & 1850,1950 & N/A & 1400,1950 & N/A & 1100,2000 & 350,2000 & 150,2000 & 0,2050\\
\hline
R & 1 & 2 & 3 & 4 & 5 & 6 & 7 & 8 & 9 & 10 & 12\\
\hline
\end{tabular}
\end{table*}
\subsection{Failure Rate}
\label{sec:evalfr}

\begin{figure}[!ht]
\begin{center}
\includegraphics[height=2.5in, width=3.4in]{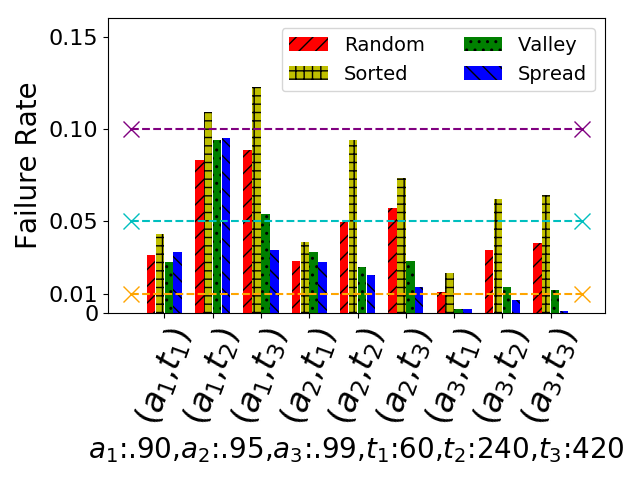}
\end{center}
\vspace*{-5mm}
\caption{Failure rate evaluation of different pilot selection algorithms}
\label{fig:algorithmfr}
\end{figure}

Figure \ref{fig:algorithmfr} shows failure rates of different pairs of target availability and lease period. As shown in the figure, Random and Sorted show the results of the same algorithms described in Section \ref{sec:exploration}. Our model, shown as Valley and Spread, can meet all target availability levels. Interestingly, as a revised algorithm of Valley, Spread shows lower failure rates than Valley. This proves that evenly distributing start time of selected pilots is an effective technique to overcome the variance in the pilot lifetime PDF.

\subsection{Redundancy}
\label{sec:evalredundancy}

\begin{figure}[!ht]
\begin{center}
\includegraphics[height=2.5in, width=3.4in]{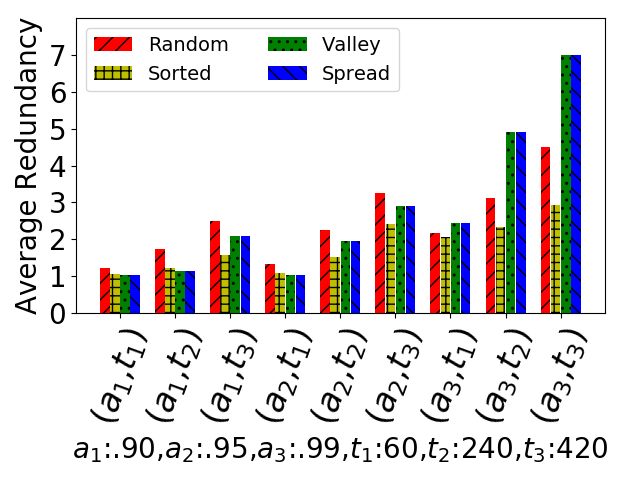}
\end{center}
\vspace*{-5mm}
\caption{Redundancy costs of different pilot selection algorithms}
\label{fig:algorithmredundancy}
\end{figure}

Figure \ref{fig:algorithmredundancy} shows the redundancy costs of different availability levels and lease periods. Unsurprisingly, the Sorted algorithm uses the minimum number of replicas. Compared with Random, Valley and Spread can use less redundancy to achieve higher availability in some test cases especially with low target availability and short lease period - (0.90, 60), (0.90, 240), (0.90, 420) and (0.95, 60).

Figure \ref{fig:algorithmredundancy} also suggests that we should always spread out the pilots based on their start time because it improves availability without introducing any redundancy overheads.

\subsection{System Utilization}
\label{sec:evalutilization}

\begin{figure}[!ht]
\begin{center}
\includegraphics[height=2.5in, width=3.4in]{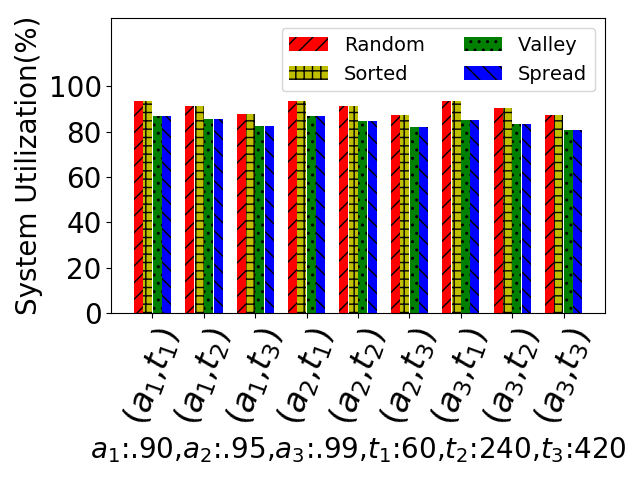}
\end{center}
\vspace*{-5mm}
\caption{System utilization of different pilot selection algorithms}
\label{fig:algorithmutilization}
\end{figure}

Figure \ref{fig:algorithmutilization} shows the pilot utilization of different selection algorithms. In general, Random and Sorted algorithms have a larger candidate pool compared with Valley and Spread algorithms. Random and Sorted are able to select all available pilots; however, Valley and Spread cannot select any pilots if available pilots are not within any pre-defined valleys. Thus the first two algorithms have higher system utilization than the latter two. As shown in Table \ref{tab:valleys}, if all pilots are close to the end of the retire time ($>$ 2100 minutes) and not enough pilots are in the valleys at a certain time point, the scheduler should not choose any pilots. As a result, these tasks should be held by the scheduler until there are more eligible pilots in the valleys later. We call such a mechanism as delay scheduling.

\subsection{Delay Scheduling - Constraining Redundancy}
\label{sec:delayscheduling}

As shown in Table \ref{tab:valleys}, a task might have to be replicated by a large number of times in certain situations. For example, given a target availability of 0.99 and a lease period of 420. The last valley that is able to cover all pilot lifetime from 0 to 2060 needs to select 12 pilots in the valley. Such an aggressive replication scheme can quickly exhaust the resources in the system. To overcome the workload burden, we propose an algorithm - delay scheduling which is able to control the upper bound of redundancy for a task. The idea is that we can define an upper bound for the largest valley in which the maximum number of replicas to replicate a task is acceptable. The scheduler stops moving to a larger valley after the maximum allowed valley fails to select enough pilots. The system has to hold the tasks in the batch queue until there are enough pilots available in the upper-bounded valley. For example, if the system put redundancy limit to 6, all valleys after v6 in Table \ref{tab:valleys} are cut off. Let us take the pair of the availability of 0.99 and the lease period of 420 as an example. If there are not enough pilots that exist between 1400 and 1950 minutes, the scheduler postpones scheduling any task until it can find 6 pilots with that valley. 

\begin{figure}[ht!]
     \centering
     \subfloat[Failure Rate]{\includegraphics[height=1.5in, width=2in]{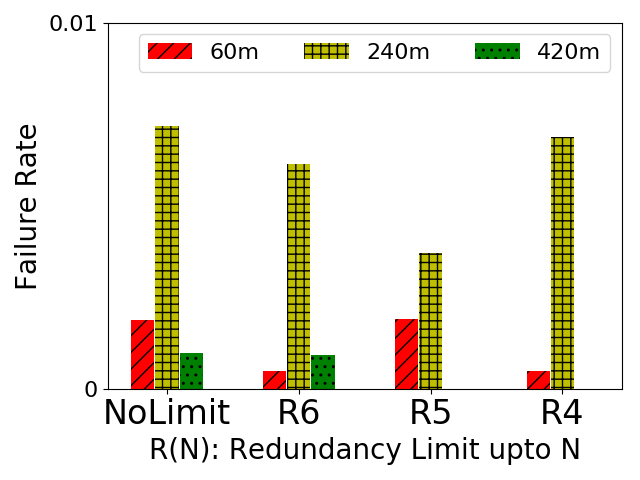}\label{fig:redundancylimitfr}}\\
     \subfloat[Redundancy]{\includegraphics[height=1.4in, width=1.7in]{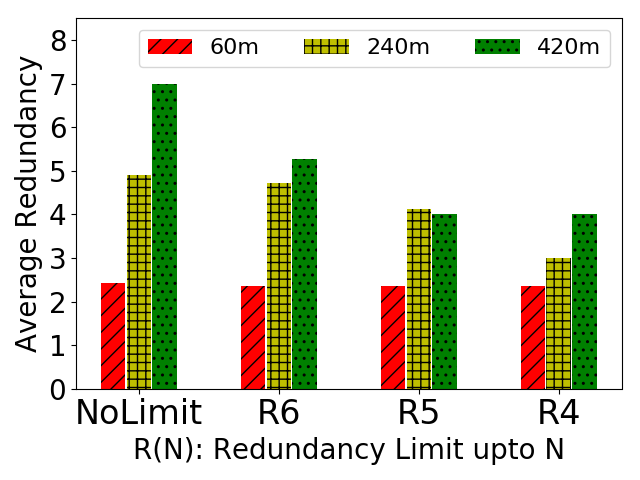}\label{fig:redundancylimitredundancy}}
     \subfloat[System Utilization]{\includegraphics[height=1.4in, width=1.7in]{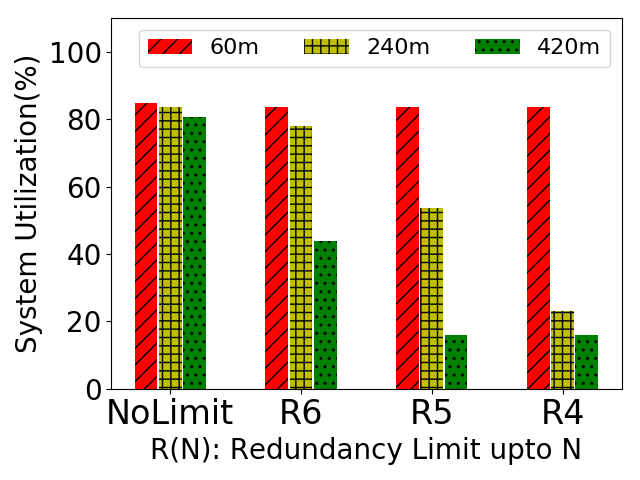}\label{fig:redundancylimitutilization}}
     \caption{Redundancy and utilization tradeoff of different redundancy upper bound. The target availability shown in the figures is 0.99.}
     \label{fig:redundancylimittradeoff}
\end{figure}

We tested the redundancy limit of 6, 5, 4. Only the larger target availability gets affected by the redundancy limit because the smaller availability does not exceed the limit. Thus, we only show the pairs of (0.99, 60), (0.99, 240) and (0.99, 420) in Figure \ref{fig:redundancylimittradeoff}. As shown in Figure \ref{fig:redundancylimitfr}, all three pairs achieve the target availability of 0.99. The lease of 420 minutes reaches almost zero failure rate on the redundancy limit 5 and 4 because it uses valleys up to [1850, 1950] minutes. In such a small pilot age range, most of the pilots are highly reliable. Figure \ref{fig:redundancylimitredundancy} shows that the redundancy keeps below the limit as expected.

Despite successfully keeping redundancy within a certain limit, delay scheduling can lower the system utilization. Let us look at the lease of 420 minutes. The scheduler only allows the pilots whose ages are in [1850, 1950]. Compared to the whole lifetime range [0, 2100], only a small portion of pilots are able to be selected. Figure \ref{fig:redundancylimitutilization} shows the system utilization on redundancy limits. As seen in the figure, only 14.5\% of pilots can go to the candidate pool. If we allow all valleys shown in Table \ref{tab:valleys} to be used in selection, 73.5\% of pilots are able to be selected in the system.
\section{Conclusions and Future Work}
\label{sec:conclusion}

In this paper, we described a new model - Trua that dynamically tunes task redundancy based on system failure rate. Our work showed the existing model fails to estimate the real-world system failure rate. The techniques in the Trua, such as anomaly detection and random selection in valleys can also be applied to other virtualized scientific grids that suffer temporal and spatial failures. Our work bridges the gap between theory and practice and provides a statistical approach to achieve user-defined availability.

In the future, we plan to enhance the Trua to an online system that allows estimating pilot lifetime and selecting pilots up to the point that prediction is made. The minimum size of a time frame that is required to accurately estimate the pilot lifetime is not analyzed in the paper. Although Section \ref{sec:framepdf} proves the pilot lifetime is stationary over a certain period of time, cross-validation among different time frames would be helpful to reduce the overfitting effects in the model.


\section*{Acknowledgment}

This work was supported by NSF award PHY-1148698, via sub-award from the University of Wisconsin-Madison. This research was done using resources provided by the Open Science Grid and the Holland Computing Center of the University of Nebraska.

\bibliographystyle{IEEEtran}
\bibliography{bib/references}

\newpage

\end{document}